\documentstyle[preprint,aps,psfig,floats]{revtex}
\begin{document}

\title{
\begin{flushright}
\vspace{-1cm}
{\normalsize IU/NTC 98-02}
\vspace{1cm}
\end{flushright}
Chiral Dynamics of Low-Energy \\
 Kaon-Baryon Interactions with Explicit Resonance}
\author{ B. Krippa$^{**}$ and J.T.Londergan}
\address{Department of Physics and Nuclear Theory
Center, Indiana University,\\
2401 Milo Sampson Lane, Bloomington, IN 47408.\\}
\maketitle
\begin{abstract}
The processes involving low energy $\overline{K}N$ and $Y\pi$ 
 interactions (where $Y= \Sigma$ or $\Lambda$) 
 are studied in the framework of heavy baryon chiral perturbation theory
 with the $\Lambda$(1405) resonance appearing as an independent field. 
 The leading and next-to-leading terms in the chiral expansion are taken 
 into account.  We show that an approach which explicitly includes the 
 $\Lambda$(1405) resonance as an elementary quantum field gives reasonable
 descriptions of both the threshold branching ratios and the energy 
 dependence of total cross sections.
%\end{abstract} 
\vskip2cm 
\noindent $^{**}${\it On leave from the Institute for Nuclear Research 
of the Russian Academy of Sciences, Moscow, 117312, Russia.} 

\end{abstract}

\newpage
\section{Introduction}

Heavy Baryon Chiral Perturbation Theory (HB$\chi$PT) applied to the 
scattering amplitudes for processes involving baryons and mesons has 
been successful in describing experimental data for a significant 
number of low-energy meson-baryon interactions \cite{Me95}. 
This method involves expansion in powers of the momentum for 
sufficiently low energies.  As the number of terms in such expansions
grows exponentially with energy, processes reasonably close to threshold 
are generally well described by this method.  However, chiral expansions 
based on HB$\chi$PT encounter a serious 
problem when they are used in systems with a resonance or bound state at 
threshold.  As such systems possess an S-matrix pole, the standard
low-order expansion techniques of chiral perturbation theory cannot
provide a satisfactory description.  Near a 
resonance the particles can in principle interact an infinite 
number of times, producing a characteristic energy dependence 
which cannot be reproduced by a power series with a small number 
of terms.

One system in which such a state appears is the NN system in the singlet
channel, which is strongly affected near threshold by the ``singlet 
deuteron'' state.  The problem of formulating a consistent chiral 
expansion for this channel has recently attracted 
much attention \cite{We92,Co97,Le97,Ka98,Bi97}. Low-energy $K^{-}p$ 
interactions are also, to a large extent, governed 
by the subthreshold  $\Lambda$(1405) resonance.
In the present paper we focus on the $\overline{K}N$ and $Y\pi$ 
systems (where $Y$ includes the $\Lambda$ and $\Sigma$ hyperons) 
at low energy.

Two methods have been suggested so far to formulate a 
chiral approach to reactions in the presence of a low-energy  
or subthreshold resonance.  In the first method \cite{We95}, one uses 
just the tree level irreducible diagrams, generated by HB$\chi$PT in 
leading and subleading order, to construct an  
effective potential.  This potential is then inserted into 
the Lippmann-Schwinger equation, which iterates the effective 
potential to all orders.  A sufficiently attractive effective 
potential will generate a resonance when iterated in this 
manner.  Using this method Kaiser, Siegel and Weise \cite{We95} 
were able to generate the $\Lambda$(1405) resonance dynamically, and 
obtained good fits to experimental data in this nonperturbative
scheme.  
 
Although iteration of an effective potential allows one to 
reproduce low-energy interactions in the vicinity of a resonance 
pole, this represents a substantial departure from HB$\chi$PT methods.  
In the standard HB$\chi$PT analysis, all possible amplitudes are
evaluated to a given order in the chiral expansion, whereas 
only ladder diagrams were summed up using the 
Lippmann-Schwinger equation.  Second, in chiral expansions one
can in principle remove the dependence of low 
energy constants on the regularization parameter by using the
renormalization group equations.  
Using the Lippmann-Schwinger equation, it is necessary to impose 
a cutoff in order to render the integral part 
of the Lippmann-Schwinger equation finite.  Moreover, it turns out
that the results of calculations depend rather strongly  
on the value of the cutoff \cite{Kr98}.  Third, as is well known in 
scattering theory, effective potentials which give identical on-shell 
results can in principle differ in their 
off-shell extrapolation, so that the results of the Lippmann-Schwinger 
calculations will also depend on the functional
form used for the effective potential.  

In the second approach, proposed by Lee, Min and Rho \cite{Le96}, 
the $\Lambda$(1405) resonance is treated 
from the beginning as an "elementary" field which participates on 
an equal footing with all other
quantum fields in the chiral Lagrangian. It was shown in 
Ref.\ \cite{Le96} that in the framework of such an approach one 
can obtain reasonable threshold branching ratios for $\overline{K}N$ 
and $Y\pi$ channels, taking into account only the leading order 
${\cal O}(Q)$ terms. However, 
in Ref. \cite{Le96} only the lowest order chiral 
Lagrangian was used in the actual calculations, and the authors 
did not compare their model with experimental data for reaction 
channels involving an initial-state $K^{-}p$ system. In the present paper 
we explore the approach suggested in Ref. 
\cite{Le96}, in which the $\Lambda$(1405) is included in the
Lagrangian as an elementary field.  We take their calculation  
one step further by including the next-to-leading order terms 
in the chiral Lagrangian, and we compare our results to experimental
reaction cross sections near threshold.

\section{Effective Chiral Lagrangian} 

The effective chiral Lagrangian can be written as a sum of terms each 
of which is proportional to a particular power of the 
characteristic momentum which we denote $Q$. Then the physical amplitude can 
be represented as a power series in $Q$. The chiral counting rules can be 
used to determine the chiral dimension
of the amplitude at any given order. The leading order terms are linear 
in $Q$, while the next-to-leading terms go as $Q^2$.  
To this order the Lagrangian we are interested in is given by 
\cite{We95}

\begin{equation}
  {\cal L}={\cal L}^{(1)}+{\cal L}^{(2)}
\end{equation}

where the leading order term is given by 

\begin{equation}
  {\cal L}^{(1)} = {i \over {8f^2}} Tr(\overline{B}[[\phi,
  \partial_0 \phi],B]) + 
  (\sqrt{2}g_{\Lambda^{*}}\overline{\Lambda^{*}}Tr((v\cdot A)B) + h.c.)
\end{equation}

 and the next-to-leading order term of the effective Lagrangian can be 
 written as follows

\begin{eqnarray}
  {\cal L}^{(2)} =  & & {i \over {2M_0}} Tr(\overline{B}((v\cdot D)^2
                                       -D^2)B)+
  b_D Tr(\overline{B}\{\chi_{+},B\}) + b_F Tr(\overline{B}[\chi_{+},B]) +           
   b_0 Tr(\overline{B}B)Tr(\chi_{+}) \nonumber \\ +                     
& & d_D Tr(\overline{B}\{A^2+(v\cdot A)^2,B\}) + 
d_F Tr(\overline{B}[A^2+(v\cdot A)^2,B]) +  
    d_0 Tr(\overline{B}B)Tr(A^2+(v\cdot A)^2) \nonumber \\+
& &  d_1 \left[ Tr(\overline{B}A_\nu)Tr(A^{\nu}B) + 
Tr(\overline{B}(v\cdot A))Tr((v\cdot A)B)\right] +\nonumber \\
& & d_2 Tr\left[\overline{B}(A_\nu BA^\nu +(v\cdot A)B(v\cdot A))\right] 
\label{secord} 
\end{eqnarray}  
We define the chiral covariant derivative $ D^\nu B=\partial^\nu B + 
[\Gamma^\nu,B] $
with a chiral connection $\Gamma^\nu$
and axial operator $A_\nu = - {1 \over {2f}} \partial_\nu \phi$
where $B$ and $\phi$ are baryon and meson field matrices in standard
form \cite{We95}. The matrix $\chi_{+}$ is given by
\begin{equation}
 \chi_{+}=- {1 \over {4f^2}}\{\phi,\{\phi,\chi\}\}
\end{equation}
 where $\chi$ is the meson mass matrix with the 
 diagonal elements $(m^{2}_\pi, m^{2}_\pi, 2m^{2}_K-m^{2}_\pi)$.
 The field $\Lambda^{*}$ describes the state $\Lambda (1405)$ introduced 
 as an elementary field with $\Lambda^{*}KN$ coupling constant 
 $g^{2}_{\Lambda^{*}}$=0.15. As in Ref. \cite{Le96} this value was 
 extracted from 
 the decay width $\Gamma$=50 MeV for the process $\Lambda (1405) 
 \rightarrow \pi\Sigma$.
 The first term in the Lagrangian ${\cal L}^{(2)}$ describes the $1/M$ 
 corrections. The form
 of this term follows directly from the corresponding relativistic 
 expression and does not contain the unknown low-energy constants. The 
 low-energy constants in front of the terms with the double derivatives 
 and mass terms are not determined by chiral symmetry and must  
 therefore be extracted from the experimental data. 
 The procedure we used to determine the constants $\{b\}$ in Eq.\ 
 \ref{secord} is similar to 
 that used in Ref. \cite{We95}. The constants $b_D$ and $b_F$ can be 
 extracted from the baryon mass splittings. Using the $SU(3)$ relations 
 for the baryon masses one can get the values $b_D = 0.06$ GeV$^{-1}$ and 
 $b_F = -0.22$ GeV$^{-1}$. The constant $b_0$ can 
 be related to the $\pi N$ sigma term. To the order we are interested in, 
 this relation is given by
 \begin{equation}
 \sigma_{\pi N}=-2 m^{2}_{\pi}(b_D+b_F+2b_0)
 \end{equation}
 In our calculations we used the value $b_0 = -0.34$ GeV$^{-1}$. The 
 uncertainty in the value of $b_0$ arises mainly from the uncertainty 
 in the pion-nucleon sigma term 
 
 The unknown effective constants $\{ d\}$ associated with the terms with 
 double derivatives in Eq.\ \ref{secord} can be extracted from 
 experimental data. 
 There are six channels ($\pi\Sigma, \pi\Lambda, \overline{K}N$)
 which can be involved in the dynamics of $K^{-}p$ interactions at low 
 energies.  There are also relations for the isospin-even $\pi N$ 
 S-wave scattering length $a^{+}_{\pi N}$
 and isospin zero kaon-nucleon S-wave scattering length $a^{0}_{KN}$, 
 which are given by the following expressions
 \begin{equation}
 4\pi (1+ {m_{\pi} \over {M_{N}}})a^{+}_{\pi N}={m^{2}_{\pi} \over {f^{2}}}
 (d_D + d_F + 2d_0 - 2b_D - 2b_F - 4b_0 -{g^{2}_{A} \over {4M_{N}}})
 \end{equation}
 and 
\begin{equation}
 4\pi (1+ {m_{K} \over {M_{N}}})a^{0}_{KN}={m^{2}_{K} \over {f^{2}}}
 ( -2d_F + 2d_0 -d_1 + 4b_F - 4b_0 -{D \over {M_{N}}}(F-D/3))
 \label{aeqn}
 \end{equation} 
 Here $g_{A}$ is the axial vector coupling constant and $F$ and $D$ are the
 SU(3) coupling constants which satisfy the relation $D+F =1.26$.  
 Eq.\ \ref{aeqn} puts constraints on the possible values 
 of the effective constants $\{ d\}$. We have used the experimental data 
 from  Refs \cite{Ko86} and \cite{Wa82}. Both scattering lengths are very 
 small: $a^{+}_{\pi N}=(-0.012 \pm 0.06)$ fm and $a^{0}_{KN} = 
 (-0.1 \pm 0.1)$ fm.
 One notes that in the expression for $a^{+}_{\pi N}$ we have not included 
 the term corresponding to the non-analytic loop correction proportional to 
 $m_{\pi}$. Since we calculate the scattering amplitudes up to order $Q^2$
 it would not be consistent to include at any point terms stemming from
 loop corrections which are of order $Q^3$.
 After imposing the constraints we are left with three undetermined 
 constants.  We fix the remaining constants using the experimental data
 describing the energy dependence of the total cross sections of 
 the reaction channels with a $K^{-}p$ system in the initial state. There 
 are six possible final channels altogether, so after fixing three 
 constant we obtain theoretical predictions for three reaction channels. 

\section{Results and Discussion}
 
First, we fix the low-energy constants $\{ d\}$ which appear in Eq.\ 
\ref{secord}. We work in 
the limit of conserved isospin and use the the following masses:  
$m_\pi$=139.6 MeV, $m_K$=493.6 MeV, $M_N$=938.3 MeV, $M_\Lambda$=1115.6 
MeV, and $M_\Sigma$=1192.5 MeV.  A set of low-energy constants
$\{ d\}$ which give a reasonable description of the experimental data 
are given in Table I.  We compare these constants with the same quantities 
defined by Kaiser {\it et al.} \cite{We95}.  These authors expand the
chiral Lagrangian to second order without an elementary $\Lambda$(1405) 
field.  The resonance is obtained by using the low-order chiral expansion
as a potential, and generating the resonance by iterating the potential
in the Lippmann-Schwinger equation.  Although the sign of each term is 
the same, the absolute values of the effective constants differ 
considerably from those obtained in Ref.\ \cite{We95}.  Thus, 
although the results of the two approaches are in good quantitative
agreement (e.g., the very similar values for the reaction cross 
sections given in Figs.\ 1-6 of this paper, and those of Kaiser 
{\it et al.}, Ref.\ \cite{We95}), the coefficients of the corresponding 
``expansion constants'' in the Lagrangian are very 
different.  The differences between expansion parameters are not 
surprising, given the different Ansatz regarding the origin of 
the $\Lambda$(1405) resonance.  If one treats the low-order chiral
expansion as a potential and iterates it in ladder approximation 
to obtain the $\Lambda$(1405), then the expansion constants of 
Ref.\ \cite{We95} are appropriate.  If one treats the $\Lambda$(1405)
as an elementary field coupled to the $\bar{K} N$ system, then 
the expansion constants of the ``residual'' system are given by
our values.  

\begin{figure}
\centering{\ \psfig{figure=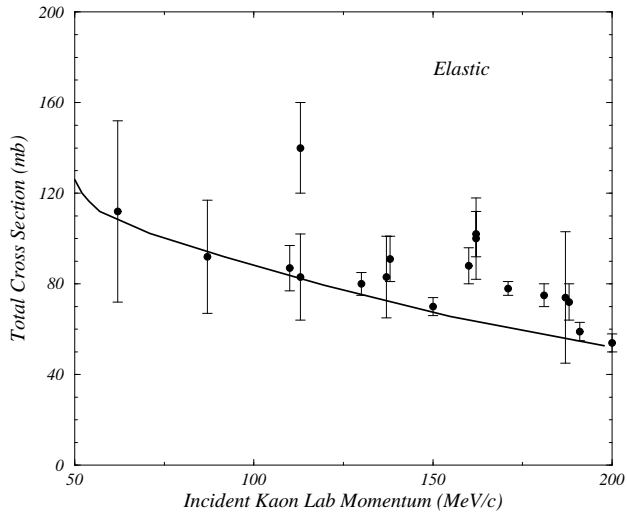,height=8cm,angle=270}}
\vskip1.5cm
\caption{Results of our theoretical calculations compared with 
experimental data \protect\cite{nowak,tovee} for $K^-p$ elastic 
scattering.}
\end{figure}

In Figs.\ 1-6, we compare the results of our calculations
of the total cross sections for all channels with the experimental data 
\cite{nowak,tovee}.  The agreement with experiment is quite reasonable,   
indicating that the leading and subleading contributions including 
an explicit $\Lambda$(1405) resonance play a 
dominant role in the low-energy dynamics of the $K^{-} p$ system. It 
is worth mentioning that without the contribution from the $\Lambda$(1405) 
resonance the probability for the process $K^{-}p \rightarrow 
\pi^{-}\Sigma^{+}$ would be suppressed at leading order. Moreover, the
cross sections of other reactions would also be much too low.  The one
case where our calculations fails to reproduce experiment is the 
$K^-p \rightarrow \pi^0 \Lambda$ channel shown in Fig.\ 5, where our 
calculated cross section is lower than the experimental data.  This is 
probably due to the fact that the $\Lambda$(1405) resonance, which
is a prominent feature for all other channels, does not
contribute to this transition.  To bring this cross section  
closer to the experimental value will probably require
calculating the loop corrections for the reaction $K^-p \rightarrow 
\pi^0 \Lambda$. So, for all reaction channels where the $\Lambda$(1405) 
resonance contributes, it  
is an important ingredient in producing a realistic 
description of the experimental data. 

\begin{figure}
\centering{\ \psfig{figure=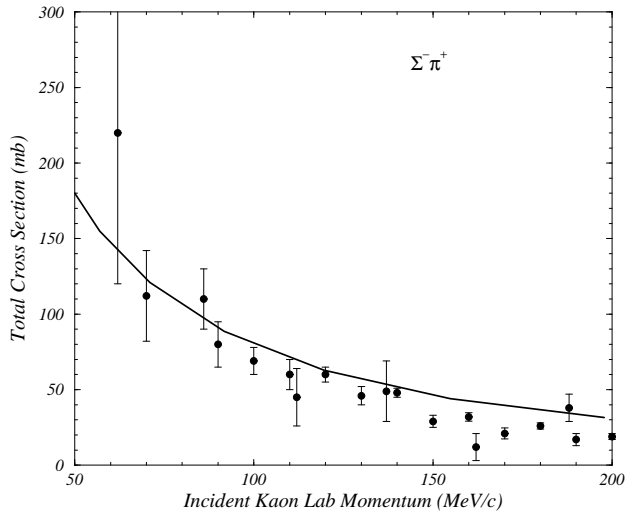,height=8cm,angle=270}}
\vskip1.5cm
\caption{Results of our theoretical calculations compared with 
experimental data \protect\cite{nowak,tovee} for the reaction $K^-p 
\rightarrow \Sigma^-\pi^+$.}
\end{figure}
 
Our predicted cross sections agree favorably with those 
calculated by Kaiser {\it et al.}, Ref.\ \cite{We95}.  They 
used the next-to-leading order contribution (not including an 
explicit $\Lambda$(1405) field) as an effective
potential, and iterated this to all orders in a Lippmann-Schwinger
equation.  By introducing an explicit $\Lambda$(1405) ``elementary'' 
field, we are able to obtain qualitatively similar results to
Kaiser {\it et al.}, except for the channel $K^- p \rightarrow 
\Lambda \pi^0$ as discussed previously.  

\begin{figure}
\centering{\ \psfig{figure=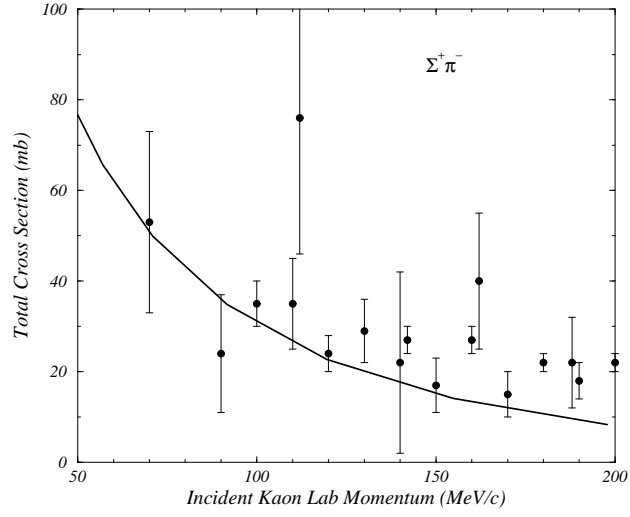,height=8cm,angle=270}}
\vskip1.5cm
\caption{Results of our theoretical calculations compared with 
experimental data \protect\cite{nowak,tovee} for the reaction $K^-p 
\rightarrow \Sigma^+ \pi^-$.}
\end{figure}
 
\begin{figure}
\centering{\ \psfig{figure=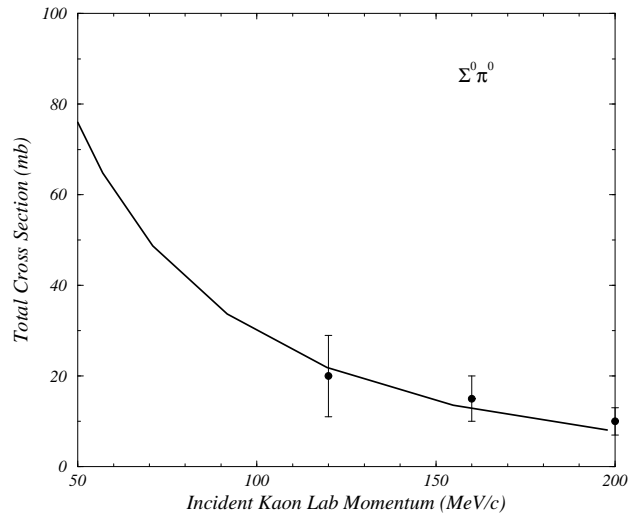,height=8cm,angle=270}}
\vskip1.5cm
\caption{Results of our theoretical calculations compared with 
experimental data \protect\cite{nowak,tovee} for the reaction $K^-p 
\rightarrow \Sigma^0\pi^0$.}
\end{figure}
 
The other major difference between the two calculations is that 
the iterated potential results are able to obtain cusps at 
inelastic thresholds, where we do not obtain cusp behavior.   
Since we are neglecting ${\cal O}(Q^3)$ corrections, we do not expect our 
calculations to be able to reproduce subtle features like threshold 
cusps.  Such cusps appear in theoretical calculations as the energy of 
the incoming kaon crosses the threshold energy of a reaction channel. 
Threshold cusps have not seen in the experimental data for $K^- p$ 
reactions.  However, the data are characterized by rather large error 
bars so that cusp singularities could very well be hidden within 
the present experimental uncertainties. Such cusps have proven 
notoriously difficult to observe in reaction cross sections.  

\begin{figure}
\centering{\ \psfig{figure=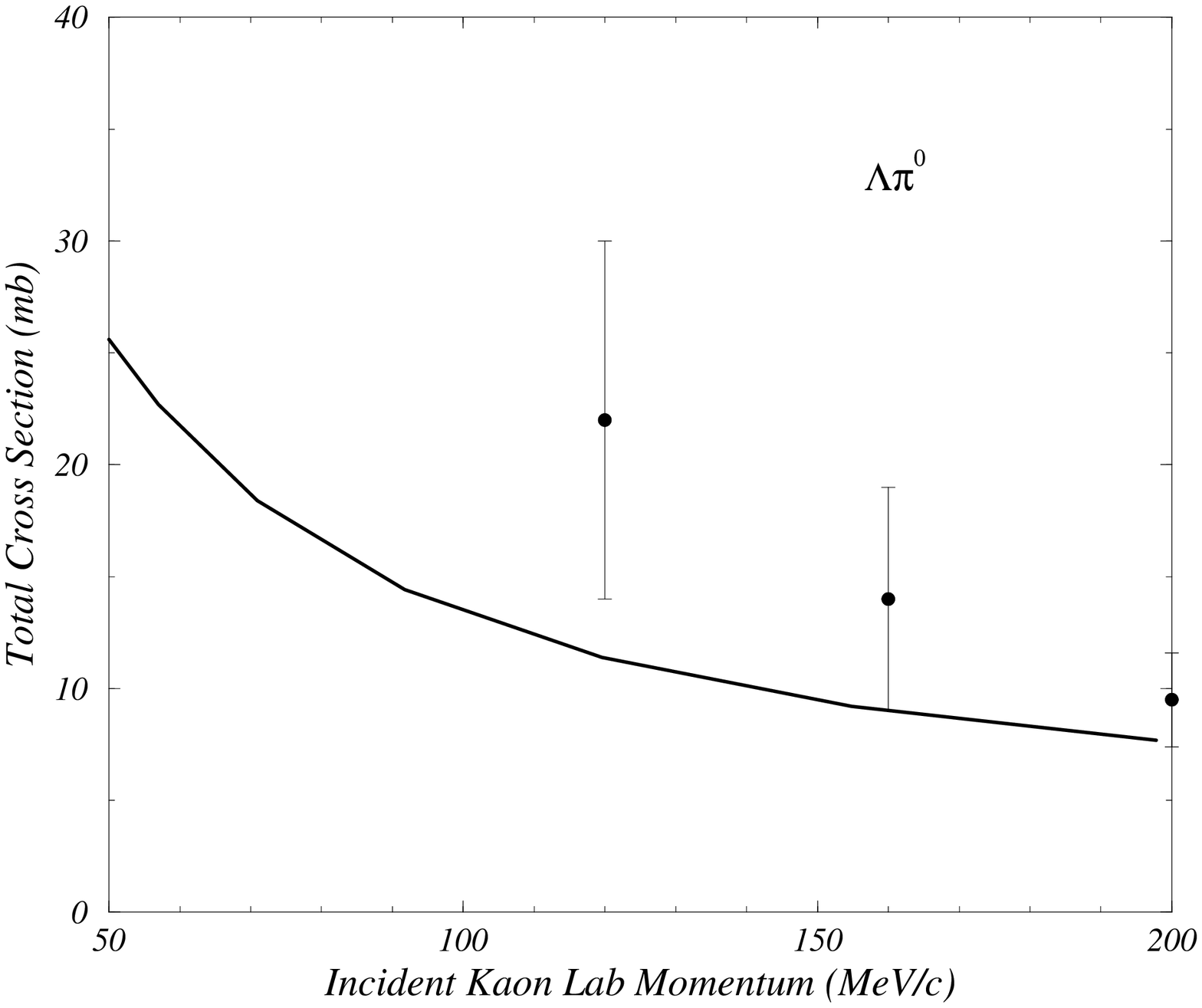,height=8cm,angle=270}}
\vskip1.5cm
\caption{Results of our theoretical calculations compared with 
experimental data \protect\cite{nowak,tovee} for the reaction $K^-p 
\rightarrow \Lambda \pi^0$.}
\end{figure}
 
\begin{figure}
\centering{\ \psfig{figure=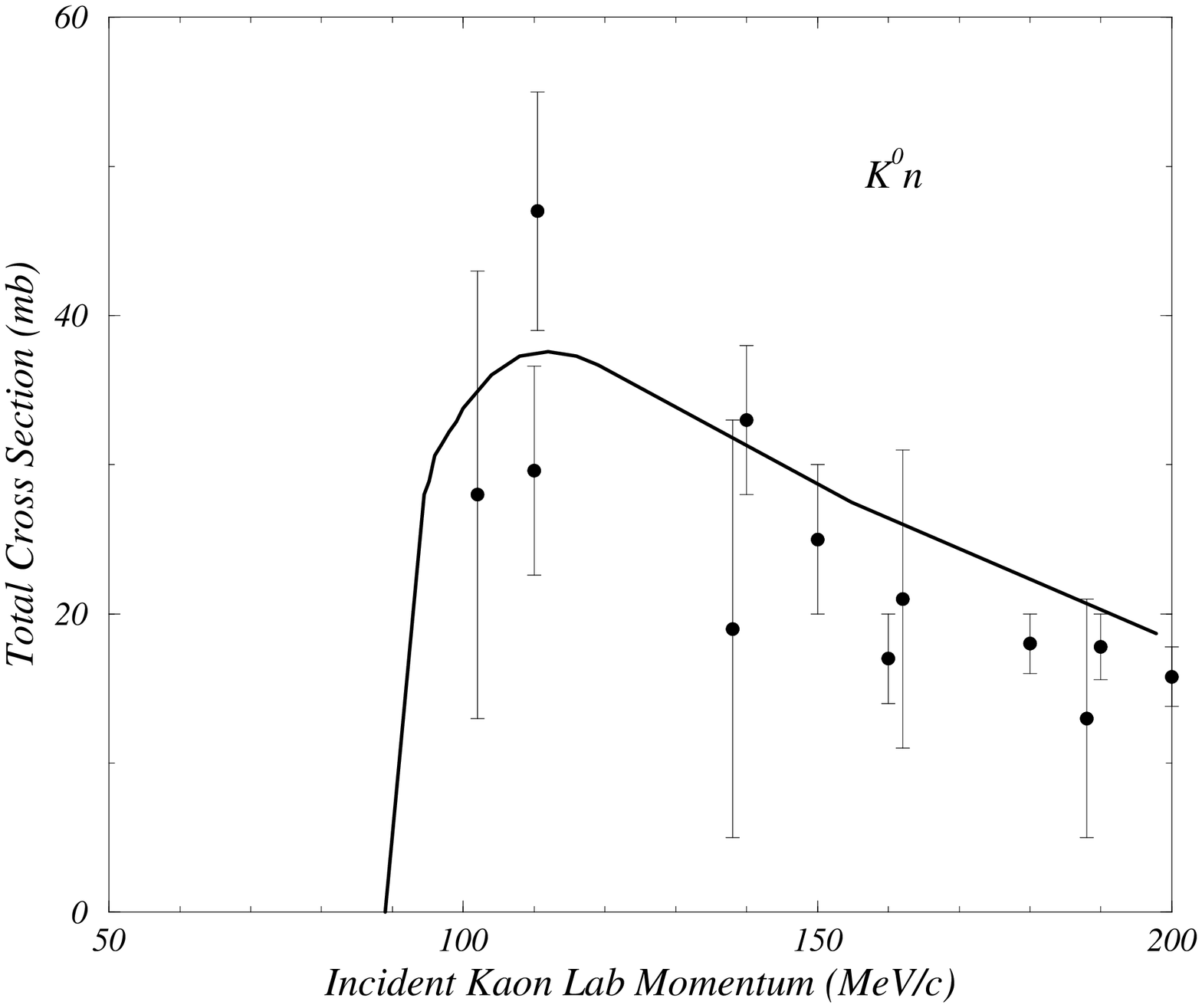,height=8cm,angle=270}}
\vskip1.5cm
\caption{Results of our theoretical calculations compared with 
experimental data \protect\cite{nowak,tovee} for the reaction $K^-p 
\rightarrow K^0 n$.}
\end{figure}
 
Next we make a few remarks concerning the loop corrections. 
They start to contribute at order ${\cal O}(Q^3)$ and are necessary  
to ensure unitarity and take into account the 
effect of coupled channels.  In addition, the $1/M^2$ and 
$SU$(3) violating effects should also be included at this order.
However, all one loop corrections in an $SU$(3) multiplet need to be 
taken into account in a consistent formalism.  At present, such a 
program would not be feasible since at this order the number of 
low-energy constants (needed to absorb divergences of the loop diagrams) 
would be much larger than the number of observables available to fix  
these constants.

In our calculation we keep the mass of the $\Lambda$(1405) resonance 
real, {\it i.e.} we treat it as a zero-width resonance.  
It would not be consistent to include the width of the $\Lambda$(1405) 
state since the width arises from loop corrections, and thus is of
higher order. We also keep the coupling constant $g_{\Lambda^{*}}$ 
the same for all channels involved.  This is an exact result in the limit
of perfect $SU$(3) symmetry. The coupling constants in various
channels may differ when we include terms of order ${\cal O}(Q^3)$.
The leading $SU$(3) violating corrections to the coupling constant 
$g_{\Lambda^{*}}$ have been computed in Ref. \cite{Sa94}. These 
corrections reduce the value of the $\overline{K}N$ 
coupling compared to $\Sigma\pi$. To evaluate how these 
corrections affect the results of calculations, one needs to include 
all possible one loop corrections.  As we noted previously, 
that is not feasible at present due to the large number
of unconstrained constants.

The other interesting physical quantities in low-energy $K^- p$ scattering 
are the threshold branching ratios
\begin{eqnarray}
    \gamma &=& {{\Gamma(K^-p \rightarrow \pi^+ \Sigma^-)} \over
       {\Gamma(K^-p \rightarrow \pi^- \Sigma^+)}} = 2.36 \pm 0.04  
       \nonumber \\
    R_c &=& {{\Gamma(K^-p \rightarrow charged ~particles)} \over
       {\Gamma(K^-p \rightarrow all)}} = 0.664 \pm 0.011  \nonumber \\
    R_n &=& {{\Gamma(K^-p \rightarrow \pi^0 \Lambda)} \over
       {\Gamma(K^-p \rightarrow ~all~ neutral~ states)}} = 0.189 \pm 0.015
\label{ratcalc}
\end{eqnarray}

The experimental values are taken from Refs.\cite{nowak}
and \cite{tovee}.   In Table II we show the theoretical results for
the branching ratios of Eq.\ \ref{ratcalc}, compared with the
experimental values.  The leading order results, of order 
${\cal O}(Q^1)$, were calculated earlier by Lee {\it et al.}, 
Ref. \cite{Le96}.  They can be compared with our results calculated
to order ${\cal O}(Q^2)$.  
Taking into account the next-to-leading corrections 
in the effective chiral Lagrangian clearly leads to
significant improvement in the theoretical description.  
The second order contributions to 
$\gamma$ and $R_c$ constitute 18\% and 48\% corrections, respectively, 
to the first-order results, and make the resulting branching
ratios in very good agreement with experiment.  Although the 
branching ratio $R_n$ is significantly increased over the lowest-order
result, it is still much lower than the experimental result. The 
discrepancy probably arises for the same reason that we fail to 
fit the $K^- p \rightarrow \pi^0 \Lambda$ cross section, namely that
the $\Lambda$(1405) resonance does not contribute to this reaction,
and that higher-order loop corrections will presumably be necessary
to reproduce this branching ratio.  
More precise measurements of all of these reactions 
would be very useful in constraining the parameters of the chiral
Lagrangians.

\section{Conclusions} 
 
We have applied the effective chiral Lagrangian to describe low-energy
interactions in channels coupled to the $K^- p$ system. Our Lagrangian 
includes the terms of leading and next-to-leading
order supplemented by a term describing the process of formation and decay
of the $\Lambda$(1405) resonance. In our approach the $\Lambda$(1405) 
state appears as an independent quantum field, on the same footing as any 
other meson or baryon field included in the 
chiral Lagrangian. The effective low-energy constants were extracted 
from the available experimental data on the different reactions 
involving a $K^- p$ channel in the
initial state.  Constraints on the expansion parameters were provided 
by fitting the isospin-even $\pi N$ S-wave scattering 
length $a^{+}_{\pi N}$ and isospin zero kaon-nucleon S-wave scattering 
length $a^{0}_{KN}$.  The $\Lambda$(1405) provides a dominant 
effect in practically every reaction channel, with the exception
of the $K^- p \rightarrow \Lambda \pi^0$ process. 

We obtained satisfactory agreement between our theoretical calculations 
and the experimental cross sections. We also analyzed three threshold 
branching ratios.  In two cases we obtained very good agreement with 
the data, and inclusion of the second-order contribution significantly
improved the calculated branching ratio.   For the branching ratio $R_n$, 
although the second-order result was twice as large as the branching 
ratio obtained in leading order, the theoretical result remained far
below experiment. The loop corrections, which are needed to take into 
account unitarity and coupled channel effects, might help
to correct the theoretical prediction for the ratio $R_n$. 
Such calculations, however, are not feasible at present because 
one has more effective constants (needed to absorb
divergences appearing in the loop integrals) than can be fixed from
experimental data.  
 
A natural extension of our formalism would be to analyze the processes 
of photo- and electroproduction of a kaon from a proton. This extension 
is clearly parameter free. Experimental data exist for the reactions 
$\gamma p \rightarrow K^+ \Lambda$ and
$\gamma p \rightarrow K^+ \Sigma$ \cite{Bo94}. An analysis of these 
experimental data would help constrain the known low energy effective 
constants, and would determine the values of previously unknown 
constants. Extending the calculations to such reactions might provide 
enough constraints that we could include loop
corrections, which are necessary to make this approach 
fully consistent.  

\section*{Acknowledgments}
One of us (B.K) would like to thank Prof. B.Holstein for stimulating 
discussions, and the Indiana University Nuclear Theory Center for the
warm hospitality extended to him during his visit.  This work was
supported in part by the National Science Foundation under contract
NSF-PHY-9722706.

\vskip 0.5truein

\begin{center}
{\Large FIGURE CAPTIONS}
\end{center}
\medskip
\begin{itemize}
\item{FIG.1} Results of our theoretical calculations compared with 
experimental data \protect\cite{nowak,tovee} for $K^-p$ elastic 
scattering. 
\item{FIG.2}  Results of our theoretical calculations compared with 
experimental data \protect\cite{nowak,tovee} for the reaction $K^-p 
\rightarrow \Sigma^-\pi^+$. 
\item{FIG.3}  Results of our theoretical calculations compared with 
experimental data \protect\cite{nowak,tovee} for the reaction $K^-p 
\rightarrow \Sigma^+ \pi^-$. 
\item{FIG.4 } Results of our theoretical calculations compared with 
experimental data \protect\cite{nowak,tovee} for the reaction $K^-p 
\rightarrow \Sigma^0\pi^0$. 
\item{FIG.5}  Results of our theoretical calculations compared with 
experimental data \protect\cite{nowak,tovee} for the reaction $K^-p 
\rightarrow \Lambda \pi^0$.
\item{FIG.6}  Results of our theoretical calculations compared with 
experimental data \protect\cite{nowak,tovee} for the reaction $K^-p 
\rightarrow K^0 n$.
\end{itemize}

\newpage

\begin{table}
\caption{Low-energy expansion constants for the chiral Lagrangian of 
Eq.\ \protect\ref{secord}. Our constants are
compared with the local and separable chiral potentials of Kaiser 
{\it et al.}, Ref.\ \protect\cite{We95}.}
\begin{tabular}{lddddd}  
  Model & $d_D$ & $d_F$ & $d_0$ & $d_1$ & $d_2$  \\ \hline
  Kaiser local & -0.02 & -0.29 & -0.66 & 0.23 & -0.39 \\ 
  Kaiser sep'l & -0.24 & -0.43 & -0.40 & 0.28 & -0.62 \\ 
  Our values & -1.70 & -0.10 & -0.45 & 0.02 & -2.3 \\ 
\end{tabular}
\label{table1}
\end{table}

\begin{table}
\caption{Threshold branching ratios in chiral perturbation expansion
with elementary $\Lambda$(1405) field. Branching ratios are defined
in Eq.\ \protect\ref{ratcalc}. Our second-order branching ratios are
compared with those obtained from the lowest-order calculation of 
Ref.\ \protect\cite{Le96}.}
\begin{tabular}{lddr}  
  Ratio & ${\cal O}(Q^1)$ & ${\cal O}(Q^2)$ & Expt. \\ \hline
  $\gamma$ & 1.93 & 2.27 & $2.36 \pm 0.04$ \\ 
  $R_c$ & 0.46 & 0.679 & $0.664 \pm 0.011$ \\ 
  $R_n$ & 0.025 & 0.051 & $0.189 \pm 0.015$ \\ 
\end{tabular}
\label{table2}
\end{table}

\newpage


\begin{thebibliography}{999}

\bibitem{Me95} For some recent reviews see: G.Ecker, Proc.\ of 
International Workshop on Hadron Physics 96: Topics on the Structure
and Interactions of Hadronic Systems, ed.\ E. Ferreira, V.L. Baltar
and T de sa Borges (World Scientific, 1997), p. 125; 
Ulf-G. Mei\ss ner, Proceedings of Second International Symposium on 
Meson-Nucleon Physics and the Structure of the Nucleon, Vancouver BC, 
Aug.\ 1997, to be published (preprint {\bf hep-ph/9707461}). 

\bibitem{We92} S.Weinberg,  Phys.\ Lett.\ {\bf B251}, 288 (1990);
 Nucl.\ Phys.\ {\bf B363}, 3 (1991).

\bibitem{Co97} S.R.Beane, T.D.Cohen and D.R.Phillips, Nucl.\ Phys.\ 
{\bf A631}, 447c (1998); preprint ({\bf nucl-th/9709062}), unpublished. 

\bibitem{Le97} G.P.Lepage, Lectures at the VIII Jorge Andre
Swieca Summer School, Brazil 1997, to be published (preprint 
{\bf nucl-th/9706029}). 

\bibitem{Ka98} D.B. Kaplan, Martin J. Savage and Mark B. Wise, 
to be published (preprint {\bf nucl-th/9801034}). 

\bibitem{Bi97} K.G.Richardson, M.C.Birse and J.A.McGovern, to 
be published (preprint {\bf hep-ph/9708435}). 

\bibitem{We95}  N.Kaiser, P.B.Siegel and W.Weise, 
Nucl.\ Phys.\ {\bf A594}, 325 (1995) 

\bibitem{Kr98} B.V.Krippa, Phys.\ Rev. {\bf C}, to be 
published (preprint {\bf hep-ph/9803332}). 

\bibitem{Le96} C.-H.Lee, D.-P.Min and M.Rho, Nucl.\ Phys.\ 
{\bf A602}, 334 (1996).

\bibitem{Ko86} R.Koch,  Nucl.\ Phys.\
{\bf A448}, 707 (1986).

\bibitem{Wa82} C.Dover and G.Walker,  Phys.\ Rep.\ {\bf 89}, 1 (1982).

\bibitem{nowak} R.J. Nowak {\it et al.},  Nucl.\ Phys.\
{\bf B139}, 61 (1978).

\bibitem{tovee} D.N. Tovee {\it et al.},  Nucl.\ Phys.\
{\bf B33}, 493 (1971).

\bibitem{Sa94} M.Savage, Phys.\ Lett.\ {\bf B331}, 411 (1994).

\bibitem{Bo94} M.Bockhorst {\it et al.},  Z.\ Phys.\
{\bf C63}, 37 (1994)

\end{thebibliography}
\end{document}